\documentclass[11pt]{amsart}
\usepackage{geometry}                
\usepackage[dvipdfmx]{graphicx}
\usepackage{amssymb}
\usepackage{epstopdf}
\usepackage{longtable}
\usepackage{threeparttable}
\usepackage{booktabs}
\usepackage{psfrag}
\usepackage{overpic}
\usepackage{color}
\usepackage{caption2}
\usepackage{subfigure}

\title{Inferring the paleo-longitude directly from the paleo-geomagnetic data}
\author{Rong Qiang Wei}
\address{College of Earth and Planet Sciences, University of Chinese Academy of Sciences, Beijing, PRC, 100049}
\email{wrq1973@ucas.edu.cn}
\date{}

\begin{document}
\maketitle
\begin{abstract}

Knowledge of the ancient geology and tectonics of the Earth owes much to paleo-magnetism, which assumes that the Earth's paleo-magnetic filed at some geo-time (or geo-time-averaged field) can be approximated as a central radial dipole at colatitude $\theta_p$ and longitude $\lambda_p$. However, it is thought that paleo-magnetism has the incapability in providing paleo-longitude. To obtain this important location parameter many other indirect methods have been developed based on different assumptions. Here we present a scanning method to derive the paleo-longitude from the usual paleo-magnetic measurements. This method takes into account the contributions to the Earth's magnetic potential from additional dipoles with their axes in the equatorial plane, which were omitted by the traditional paleo-magnetism. In this method, firstly we assume that $\theta_p$ and $\lambda_p$ are accurate (or determined well enough), and define a cost function; And secondly we minimize this function by systematically searching through all longitudes and latitudes in their domain; Finally when a local minima of this cost function reaches, the corresponding longitude is the paleo-longitude that we look for. Simultaneously the paleo-latitude is obtained. Synthetic experiments show that this method works very well when there are no errors in the geomagntic measurements (Components of magnetic field: $B_x, B_y, B_z$, or declination $D$ and inclination $I$). If there exist errors in geomagntic measurements, we recommend adding a Tikhonov regularization factor to the cost function for deriving reasonable paleo-longitude, and provide two examples. Error analysis shows that the main error sources for paleo-longitude are $B_y$ and/or $I$ in our method. In addition, such a cost function and its like could be used as a theoretical framework that can directly invert the paleo-longitude, paleo-latitude, and even the location of the paleo-geomagnetic poles simultaneously from the paleo-geomagnetic measurements through any appropriate inversion method.

\end{abstract}

{\hspace{2.2em}\small Keywords:}
paleo-longitude; paleo-magnetism; plate tectonic reconstructions; 

\hspace{6.5em} Tikhonov regularization
{\hspace{2.2em}\tiny }

\section{Introduction}\label{intro}

Obtaining quantitatively the paleo-position of continents is essential to the plate tectonic reconstructions.  The paleo-position of the continents includes basically the paleo-longitude $\lambda$ and the paleo-latitude $\bar{\theta}$ in the past. The $\bar{\theta}$ of the continents can be traditionally inferred from the paleo-magnetic data of inclination $I$ (eg., Turcotte and Schubert, 2014). This inferring is based on the assumption that the Earth's paleo-magnetic filed at some geo-time (or geo-time-averaged field) can be approximated as a dipole field. Such a dipole field can be modeled through a Gaussian spherical harmonic expansion with $n=1$ and the Gauss coefficient $g_1^1$ and $h_1^1$  equal to zero (see details in the section \ref{sec2}). In this case,  the paleo-magnetic field is axis symmetric and can not provide any information on the paleo-longitude. 

To obtain the paleo-longitude of the continents, many methods other than paleo-magnetism were developed, and different reference frames were constructed. Some authors established the hot spots absolute plate motion reference frame (e.g., M$\ddot{u}$ller et al., 1993; O' Neill et al., 2005; Torsvik et al., 2008a; Doubrovine et al., 2012), for the motion of the lithospheric plates may be reflected by the track geometry of the hot spots (Morgan, 1971).  A global hybrid reference frame, by correlating large igneous provinces and deep mantle heterogeneities at the core-mantle boundary, was established by Torsvik et al. (2008b). This reference frame assumed zero longitudinal motion of Africa before 100 Ma. By modeling of plume motions, these reference frames provided compatible reconstructions of plates with geologic and geophysical data (e.g., Doubrovine et al., 2012).  Besides, van der Meer et al. (2010) linked the lower mantle slab remnants with the global orogenic belts reconstructions, and established a sinking slab remnant reference frame. This reference frame assumed a vertical slab sinking at an average rate of $12\pm3$mm/yr.

On the other hand, researchers attempted to estimate paleo-longitude from the data associated with paleo-magnetism, especially the data of polar wander path (PWP).  For example, Mitchell et al. (2012) traced the moving trajectory of supercontinents centers in the deep geologic history and presented a true PWP derived reference frame, in which they assumed the geoid highs are stable. Wu and Kravchinsky (2014) and Wu et al. (2015) presented a synthesized method to derive paleo-longitude by geometrically parametrizing apparent PWP. The method restores the absolute motion history for the reference geometries from the Euler parameters extracted from the apparent PWPs. In this method, a paleo-colatitude correction to the reconstructions was introduced in order to keep the restored paleo-latitudes compatible with the paleo-magnetic prediction. 

Although based on different assumptions, these work above are helpful to estimate the paleo-position of the continents, and to understand how the fragments of the outer shell of the Earth have moved relative to a reference system over geological timescales. Here we present another alternative approach which we call it scanning method to infer the paleo-longitudes (and simultaneously the paleo-latitudes) from the usual paleo-magnetic measurements. The related theory will be given in section \ref{sec2}. Because inferring the paleo-longitude based on this theory is a nonlinear problem, we present a simple approach to solve it. Then we test this method with synthesized data, and discuss the error from different sources in the section \ref{sec3}. Finally we give a short discussion in the section \ref{sec4} and draw some conclusions in the section \ref{sec5}.

 \section{Theory and Methodology}\label{sec2}

\subsection{Theory}

 We start from the well-known Gauss's spherical expression for the potential of the geomagnetic field,

 \begin{equation}\label{eq1}
 V=a\sum_{n=1}^{\infty}\sum_{m=0}^{n}(\frac{a}{r})^{n+1}P_n^m(\theta)(g_n^m\cos m\lambda+h_n^m\sin m\lambda)
 \end{equation}
where $a$ is the Earth's radius, $P_n^m(\theta)$ Schmidt polynomials which are related to the associated Legendre polynomials, $g_n^m$ and $h_n^m$ Gauss coefficients of order $n$ and degree $m$, $r$ the distance from the Earth's center, $\theta$ colatitude and $\lambda$ longitude. 

\begin{figure}[htp]
\includegraphics[scale=0.8,bb=46 266 540 486]{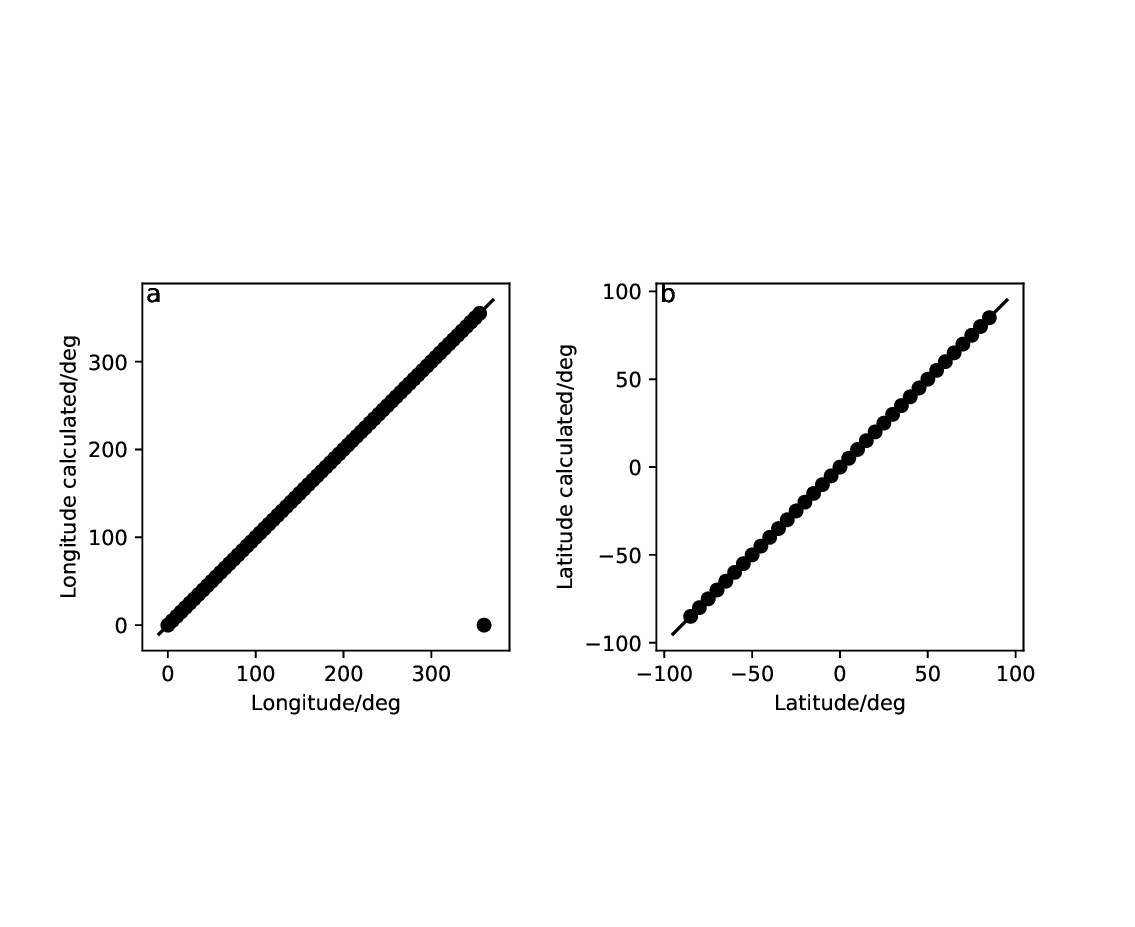}
\renewcommand{\figurename}{Fig.}
\caption{\footnotesize{a. Longitudes derived from Eq. (\ref{eq6}) vs. the set longitudes. The latitude is fixed at $39^\circ$N.  b. Latitudes derived from Eq. (\ref{eq6}) vs. the set latitudes. The longitude is fixed at $120^\circ$E. $\widetilde{B_x}$, $\widetilde{B_y}$, $\widetilde{B_z}$ are from International Geomagnetic Reference Field (IGRF) model in the year 2009.}}
\label{fig1}
\end{figure}

Generally $V_1=V\vert_{n=1}$  is taken as the potential of the centered dipole field, which is a first order approximation but the most important part of the geomagnetic field.

\begin{equation}\label{eq2}
V_1=\frac{a^3}{r^2}(g_1^0\cos\theta+g_1^1\sin\theta\cos\lambda+h_1^1\sin\theta\sin\lambda)
\end{equation}
where the term $g_1^0$ is the strongest component of the field. It describes a magnetic dipole at the center of the Earth and
aligned with the Earth's rotation axis. The terms $g_1^1$ and $h_1^1$ are the next strongest parts. They describe contributions to the magnetic potential from additional dipoles with their axes in the equatorial plane. 

One can obtain the components of the geomagnetic dipole field $B_x, B_y, B_z$ at the surface as the following,

\begin{equation}\label{eq3}
\left\{
\begin{array}{lll}
 B_x=\frac{1}{r}\frac{\partial V}{\partial \theta}\vert_{r=a}&\approx \frac{1}{r}\frac{\partial V_1}{\partial \theta}\vert_{r=a}&=-[g_1^0\sin\theta-(g_1^1\cos\lambda+h_1^1\sin\lambda)\cos\theta]   \\
 B_y=-\frac{1}{r\sin\theta}\frac{\partial V}{\partial \lambda}\vert_{r=a}&\approx -\frac{1}{r\sin\theta}\frac{\partial V_1}{\partial \lambda}\vert_{r=a}&= g_1^1\sin\lambda-h_1^1\cos\lambda  \\
 B_z=\frac{\partial V}{\partial r}\vert_{r=a}&\approx \frac{\partial V_1}{\partial r}\vert_{r=a}&=-2[g_1^0\cos\theta+(g_1^1\cos\lambda+h_1^1\sin\lambda)\sin\theta]
\end{array}
\right.
\end{equation}

From Eq. (\ref{eq3}), the longitude $\lambda$ can be possibly inferred from the measured $B_x$, $B_y$, $B_z$ if $g_1^0$, $g_1^1$ and $h_1^1$ are known. However, $g_1^0$, $g_1^1$ and $h_1^1$ are generally difficult to be obtained. If let $g_1^1= h_1^1=0$ in Eq. (\ref{eq3}), we can immediately get $B_y=0$, and

\begin{equation}\label{eq4}
\tan I=\frac{B_z}{\sqrt{B_x^2+B_y^2}}=\frac{B_z}{B_x}=2\cot\theta
\end{equation}
where $I$ is the magnetic inclination. Eq. (\ref{eq4}) is the foundational equation of the usual paleo-magnetism. It can be seen that only $\theta$ (then the latitude) appears. 

\begin{figure}[htb]
\setlength{\belowcaptionskip}{0pt}
\centering
\begin{overpic}[scale=0.5,bb=1 133 583 650]{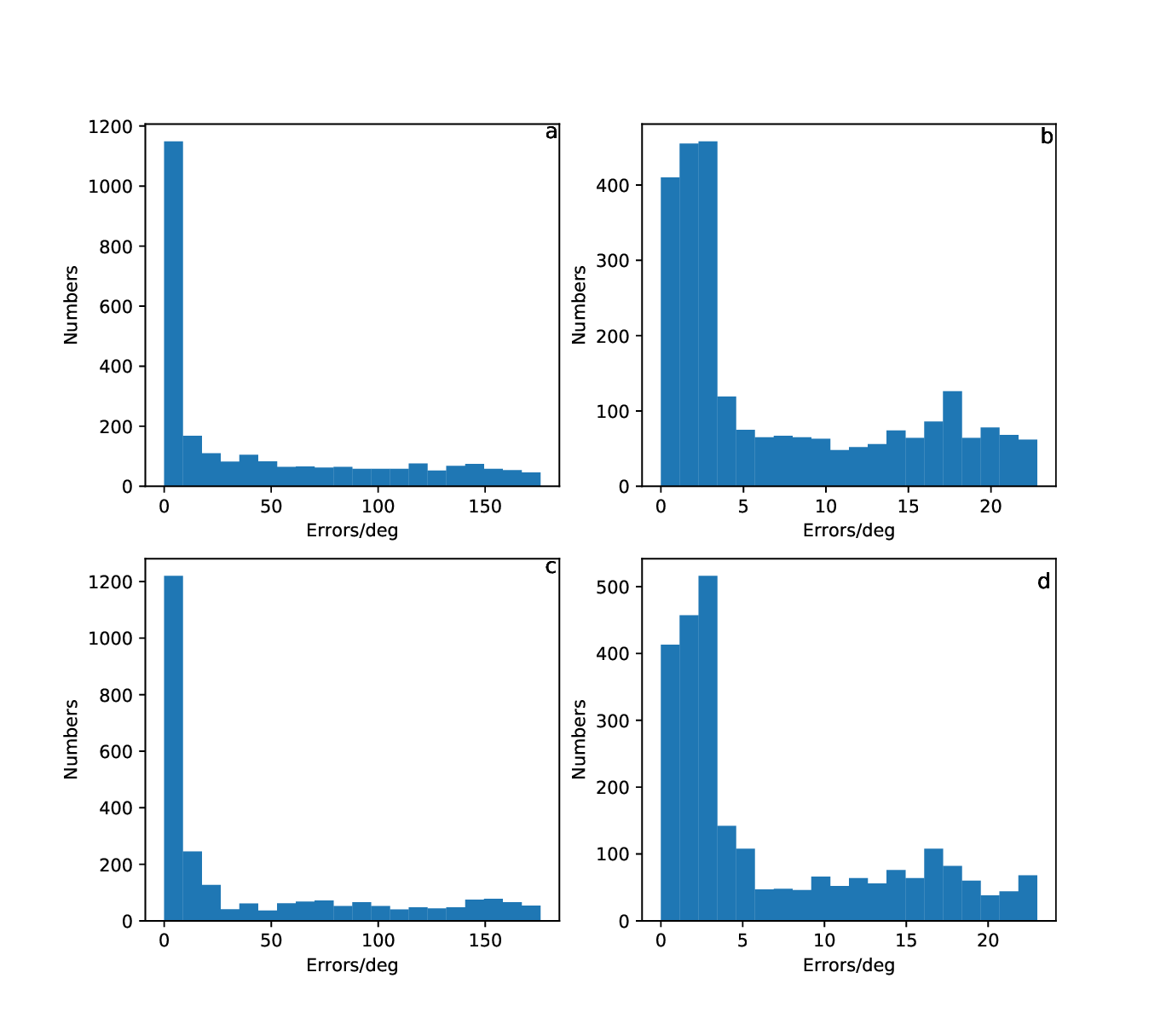}
\end{overpic}
\renewcommand{\figurename}{Fig.}
\caption{\footnotesize{a. Histogram of $\vert\lambda-\lambda_{\rm s}\vert$ (where $\lambda_{\rm s}$ is the set longitude) when an error of $0.1\widetilde{B_x}$ is added to $\widetilde{B_x}$. b. Histogram of $\vert\bar{\theta}-\bar{\theta_{\rm s}}\vert$ ($\bar{\theta_{\rm s}}$ is the set latitude) when an error of $0.1\widetilde{B_x}$ is added to $\widetilde{B_x}$. c. Histogram of $\vert\lambda-\lambda_{\rm s}\vert$ when an error of $-0.1\widetilde{B_x}$ is added to $\widetilde{B_x}$. d. Histogram of $\vert\bar{\theta}-\bar{\theta_{\rm s}}\vert$ when an error of $-0.1\widetilde{B_x}$ is added to $\widetilde{B_x}$. The $\widetilde{B_x}$, $\widetilde{B_y}$, and $\widetilde{B_z}$ for 2555 points on a longitude/latitude grid ($0.0^\circ:5.0^\circ:360.0^\circ{\rm E}, -85.0^\circ:5.0^\circ:85.0^\circ{\rm N}$) are calculated from IGRF model. The same below.}}
\label{fig2}
\end{figure}

Clearly $g_1^1$ and $h_1^1$ can not be omitted in Eq. (\ref{eq3}) to infer $\lambda$. $g_1^1$ and $h_1^1$ will be taken into account in the following approach we adopt. According to the central dipole model (e.g., Hurwitz, 1960; Alldredge and Hurwitz, 1964; Lanza and Meloni, 2006), Eq. (\ref{eq3}) can be rewritten as,

\begin{equation}\label{eq5}
 \left\{
 \begin{array}{ll}
 B_x&=-K_p[\cos\theta_p\sin\theta-\sin\theta_p\cos\theta\cos(\lambda-\lambda_p)]\\
 B_y&=K_p\sin\theta_p\sin(\lambda-\lambda_p)\\
 B_z&=-2K_p[\cos\theta_p\cos\theta+\sin\theta_p\sin\theta\cos(\lambda-\lambda_p)]
 \end{array}
 \right.
 \end{equation}
where $K_p=M_p/a^3$, and $M_p$ is the magnetic moment for the central radial dipole at colatitude $\theta_p$ and east longitude $\lambda_p$. It is easy to demonstrate that $g_1^0=-K_p\cos\theta_p$, $g_1^1=-K_p\sin\theta_p\cos\lambda_p$, $h_1^1=-K_p\sin\theta_p\sin\lambda_p$, and $B_x^2+B_y^2+(B_z/2)^2=K_p^2$. Therefore, Eq. (\ref{eq5}) is equivalent to Eq. (\ref{eq3}), and $g_1^1$ and $h_1^1$ are naturally included in the Eq. (\ref{eq5}).

From Eq. (\ref{eq5}) , the longitude $\lambda$, even $\theta$, and $K_p$,  can be possibly inferred if $B_x$, $B_y$, $B_z$, $\theta_p$, and $\lambda_p$ are known, because these quantities can be measured or estimated relatively easily. 

\subsection{Methodology}

 It is not easy to infer $\lambda$ and $\theta$ simultaneously from Eq. (\ref{eq5}), because it is a nonlinear and overdetermined problem. However, by systematically searching through all $\lambda\in [0^\circ,360^\circ]$ and $\bar{\theta}\in[-90^\circ,90^\circ]$ ($\theta=\pi/2-\bar{\theta}$) for the local minima of a cost function $R$ in Eq. (\ref{eq6}) or Eq. (\ref{eq7}), we can get $\lambda$ and $\theta$ simultaneously with the known $B_x$, $B_y$, $B_z$, $\theta_p$, and $\lambda_p$. We call this approach a scanning method because we will scan the domain of $\lambda$ and $\bar{\theta}$ for our purpose. Here, $\theta_p$ and $\lambda_p$ are assumed to have been determined well, since there has been a lot of work to determine $\theta_p$ and $\lambda_p$, and a large amount of reasonable data on them has been accumulated (eg., Torsvik et al., 2008a).
 
\begin{equation}\label{eq6}
R=\left(\frac{\widetilde{B_y}}{\widetilde{B_x}}-\frac{B_y}{B_x}\right)^2+\left[\frac{\widetilde{B_z}}{(\widetilde{B_x}^2+\widetilde{B_y}^2)^{1/2}}-\frac{B_z}{(B_x^2+B_y^2)^{1/2}}\right]^2
\end{equation}
where $B_x$, $B_y$, $B_z$ are from Eq. (\ref{eq5}) for the given $\lambda$ and $\theta$, and $\widetilde{B_x}$, $\widetilde{B_y}$, $\widetilde{B_z}$ are from measurements.

\begin{equation}\label{eq7}
R=(\widetilde{D}-D)^2+(\widetilde{I}-I)^2
\end{equation}
where $\widetilde{D}=\tan^{-1}(\widetilde{B_y}/\widetilde{B_x})$, $\widetilde{I}=\tan^{-1}\left[\widetilde{B_z}/(\widetilde{B_x}^2+\widetilde{B_y}^2)^{1/2}\right]$, $D=\tan^{-1}(B_y/B_x)$ and $I=\tan^{-1}\left[B_z/(B_x^2+B_y^2)^{1/2}\right]$.

This approach has at least two obvious advantages.  One is that we do not provide initial values for $\lambda$ and $\theta$.  The other is that the errors in $\widetilde{B_x}$, $\widetilde{B_y}$ and $\widetilde{B_z}$ (or $\widetilde{D}$ and $\widetilde{I}$) can be partially reduced. When the errors of $\Delta\widetilde{B_x}$, $\Delta\widetilde{B_y}$ and $\Delta\widetilde{B_z}$ are the same and proportional to $\widetilde{B_x}$, $\widetilde{B_y}$ and $\widetilde{B_z}$, they can be completely eliminated.  This is important in paleo-magnetic study, because usually the remanent magnetization $J_r$ of the rocks is assumed to be proportional to $B$ as the following (Eq. (\ref{eq8})), 

\begin{equation}\label{eq8}
\left\{
\begin{array}{l}
J_{rx} \approx k\frac{B_x}{\mu_0}=k' B_x\\
J_{ry} \approx k\frac{B_y}{\mu_0}=k' B_y\\
J_{rz} \approx k\frac{B_z}{\mu_0}=k' B_z
\end{array}
\right.
\end{equation}
where $\mu_0$ is permeability constant ($=4\pi\times 10^{-7}\rm{NA}^{-2}$), $k$ the proportionality constant.

\begin{figure}[htb]
\setlength{\belowcaptionskip}{0pt}
\centering
\begin{overpic}[scale=0.5,bb=1 133 583 650]{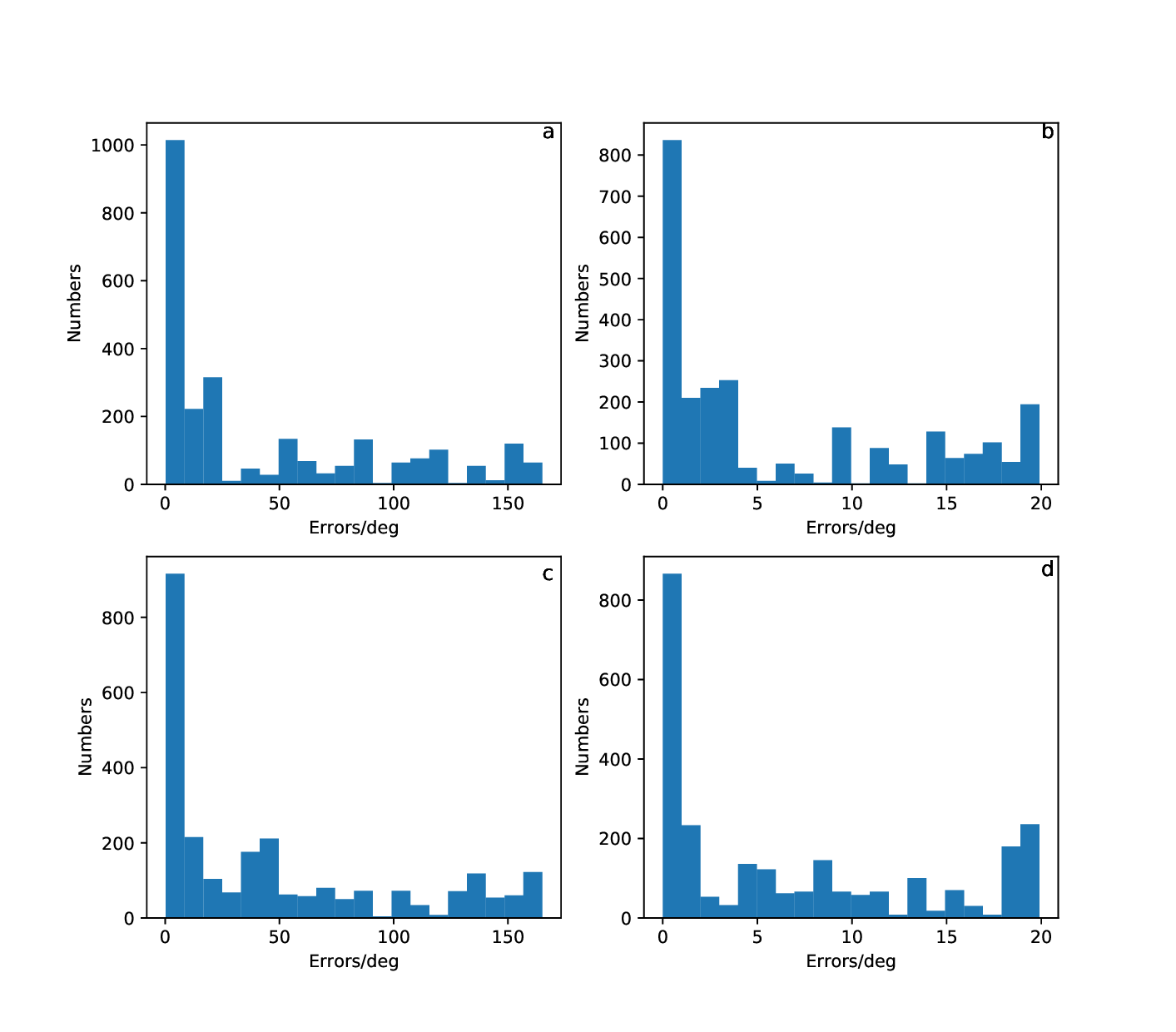}
\end{overpic}
\renewcommand{\figurename}{Fig.}
\caption{\footnotesize{a. Histogram of $\vert\lambda-\lambda_{\rm s}\vert$ when an error of $0.1\widetilde{B_y}$ is added to $\widetilde{B_y}$. b. Histogram of $\vert\bar{\theta}-\bar{\theta_{\rm s}}\vert$ when an error of $0.1\widetilde{B_y}$ is added to $\widetilde{B_y}$. c. Histogram of $\vert\lambda-\lambda_{\rm s}\vert$ when an error of $-0.1\widetilde{B_y}$ is added to $\widetilde{B_y}$. d. Histogram of $\vert\bar{\theta}-\bar{\theta_{\rm s}}\vert$ when an error of $-0.1\widetilde{B_y}$ is added to $\widetilde{B_y}$.}}
\label{fig3}
\end{figure}

\begin{figure}[htb]
\setlength{\belowcaptionskip}{0pt}
\centering
\begin{overpic}[scale=0.5,bb=1 133 583 650]{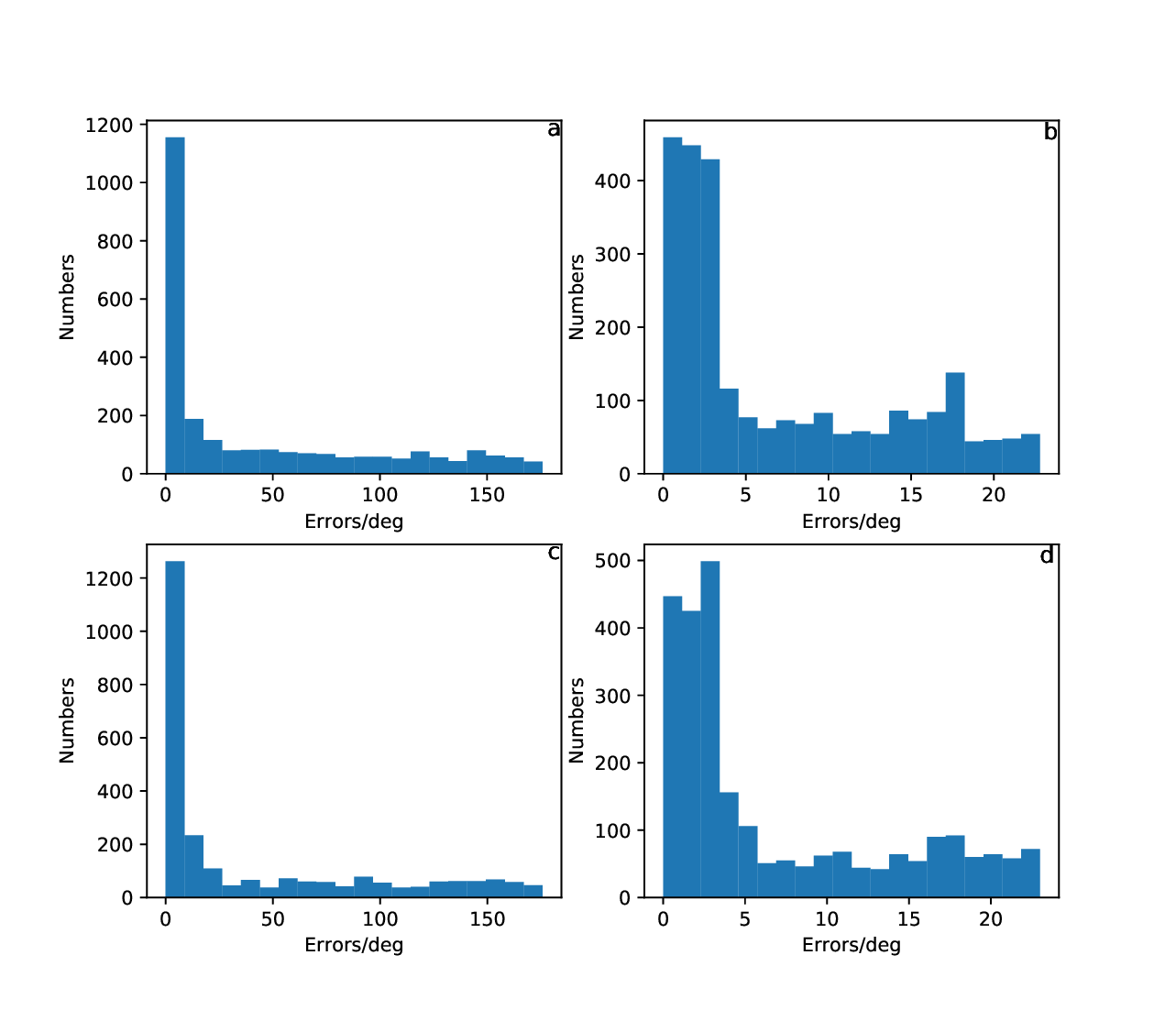}
\end{overpic}
\renewcommand{\figurename}{Fig.}
\caption{\footnotesize{a. Histogram of $\vert\lambda-\lambda_{\rm s}\vert$ when an error of $0.1\widetilde{B_z}$ is added to $\widetilde{B_z}$. b. Histogram of $\vert\bar{\theta}-\bar{\theta_{\rm s}}\vert$ when an error of $0.1\widetilde{B_z}$ is added to $\widetilde{B_z}$. c. Histogram of $\vert\lambda-\lambda_{\rm s}\vert$ when an error of $-0.1\widetilde{B_z}$ is added to $\widetilde{B_z}$. d. Histogram of $\vert\bar{\theta}-\bar{\theta_{\rm s}}\vert$ when an error of $-0.1\widetilde{B_z}$ is added to $\widetilde{B_z}$.}}
\label{fig4}
\end{figure}

Generally there are different errors in $\widetilde{B_x}$, $\widetilde{B_y}$ and $\widetilde{B_z}$, or $\widetilde{D}$ and $\widetilde{I}$. These errors may cause unstable solutions to $\lambda$ and $\theta$, and we will see this in the following synthetic experiments.  In these cases, an approach like the Tikhonov regularization should be used. Namely, we should scan all $\lambda$ and $\bar{\theta}$ in their domain for the local minima of the cost function $R$ in Eq. (\ref{eq9}) or Eq. (\ref{eq10}),

\begin{equation}\label{eq9}
R=\left(\frac{\widetilde{B_y}}{\widetilde{B_x}}-\frac{B_y}{B_x}\right)^2+\left[\frac{\widetilde{B_z}}{(\widetilde{B_x}^2+\widetilde{B_y}^2)^{1/2}}-\frac{B_z}{(B_x^2+B_y^2)^{1/2}}\right]^2+\alpha\left[(\lambda-\lambda_0)^2+(\bar{\theta}-\bar{\theta_0})^2\right]
\end{equation}
where $\alpha$ is a regularization parameter. $\lambda_0$, and $\bar{\theta_0}$ are initial values for $\lambda$ and $\bar{\theta}$, respectively.

\begin{equation}\label{eq10}
R=(\widetilde{D}-D)^2+(\widetilde{I}-I)^2+\alpha\left[(\lambda-\lambda_0)^2+(\bar{\theta}-\bar{\theta_0})^2\right]
\end{equation}

\section{Synthetic experiments}\label{sec3}
\subsection{In the case of no errors}

We test our approach with a series of synthetic experiments. The synthetic data, namely,  the components of the Earth's magnetic field at the surface, $\widetilde{B_x}$,$\widetilde{B_y}$,$\widetilde{B_z}$ for Eq. (\ref{eq6}), and $\widetilde{D}$, $\widetilde{I}$ for Eq. (\ref{eq7}), are calculated from International Geomagnetic Reference Field (IGRF) model (International Association of Geomagnetism and Aeronomy, Working Group V-MOD, 2010) when $n=1$, respectively. The time is also given randomly in the year 2009. The set longitudes and latitudes are at an interval of $5^\circ$. Both of the search step for $\lambda$ and $\bar{\theta}$ are $0.1^\circ$.  $\lambda_p$ and $\theta_p$ are calculated from Gaussian coefficients $g_1^0, g_1^1, h_1^1$ with the following Eq. (\ref{eq11}) (eg., Lanza and Meloni, 2006),

\begin{equation}\label{eq11}
\left\{
\begin{array}{ll}
\lambda_p&=\tan^{-1}(\frac{h_1^1}{g_1^1})\\
\theta_p&=\cot^{-1}(\frac{g_1^0}{\sqrt{(g_1^1)^2+(h_1^1)^2}})
\end{array}
\right.
\end{equation}

Eq. (\ref{eq11}) gives that $\lambda_p=107.87^\circ$ and $\theta_p=-79.96^\circ$ in the year 2009.

When there are no errors in $\widetilde{B_x}$, $\widetilde{B_y}$ and $\widetilde{B_z}$ (and then $\widetilde{D}$ and $\widetilde{I}$), the set $\lambda$ and $\bar{\theta}$ are inferred perfectly. Figure \ref{fig1} gives two examples.  Figure \ref{fig1}a shows the comparison of the longitudes derived from Eq. (\ref{eq6}) with the set (latitude is fixed at $39^\circ$N); Figure \ref{fig1}b shows the comparison of the latitudes derived with the set (longitude is fixed at $120^\circ$E). It can be seen that the set longitudes or latitudes are derived correctly. It should be noted that  the coordinates of the point in the lower right corner of Figure \ref{fig1}a are ($360^\circ,0^\circ$). Therefore, the longitude derived is the same to the set. 

\begin{figure}[htb]
\setlength{\belowcaptionskip}{0pt}
\centering
\begin{overpic}[scale=0.5,bb=-20 133 620 650]{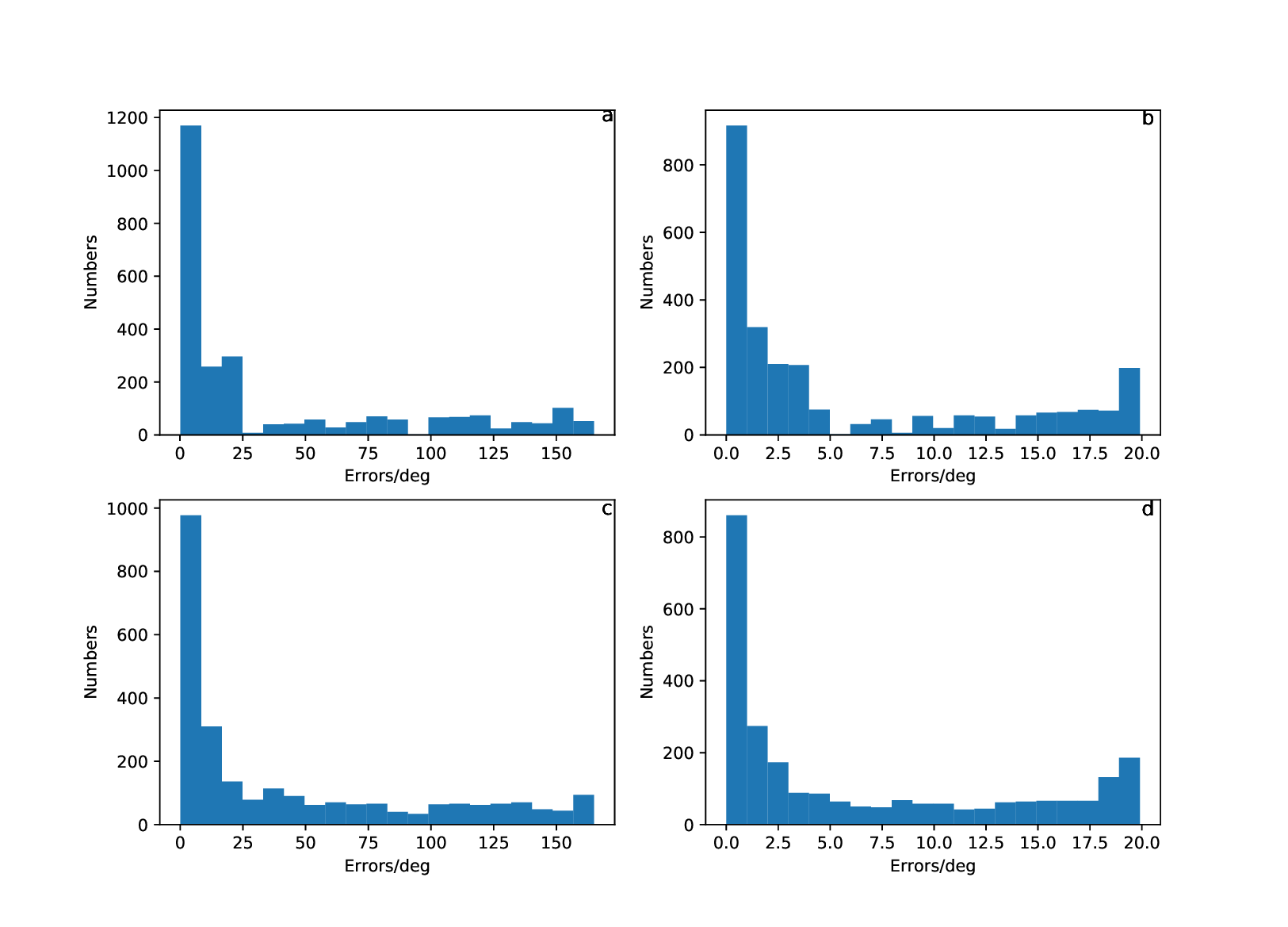}
\end{overpic}
\renewcommand{\figurename}{Fig.}
\caption{\footnotesize{a. Histogram of $\vert\lambda-\lambda_{\rm s}\vert$ when an error of $0.1\widetilde{D}$ is added to $\widetilde{D}$. b. Histogram of $\vert\bar{\theta}-\bar{\theta_{\rm s}}\vert$ when an error of $0.1\widetilde{D}$ is added to $\widetilde{D}$. c. Histogram of $\vert\lambda-\lambda_{\rm s}\vert$ when an error of $-0.1\widetilde{D}$ is added to $\widetilde{D}$. d. Histogram of $\vert\bar{\theta}-\bar{\theta_{\rm s}}\vert$ when an error of $-0.1\widetilde{D}$ is added to $\widetilde{D}$.}}
\label{fig5}
\end{figure}

\begin{figure}[htb]
\setlength{\belowcaptionskip}{0pt}
\centering
\begin{overpic}[scale=0.5,bb=-10 133 600 650]{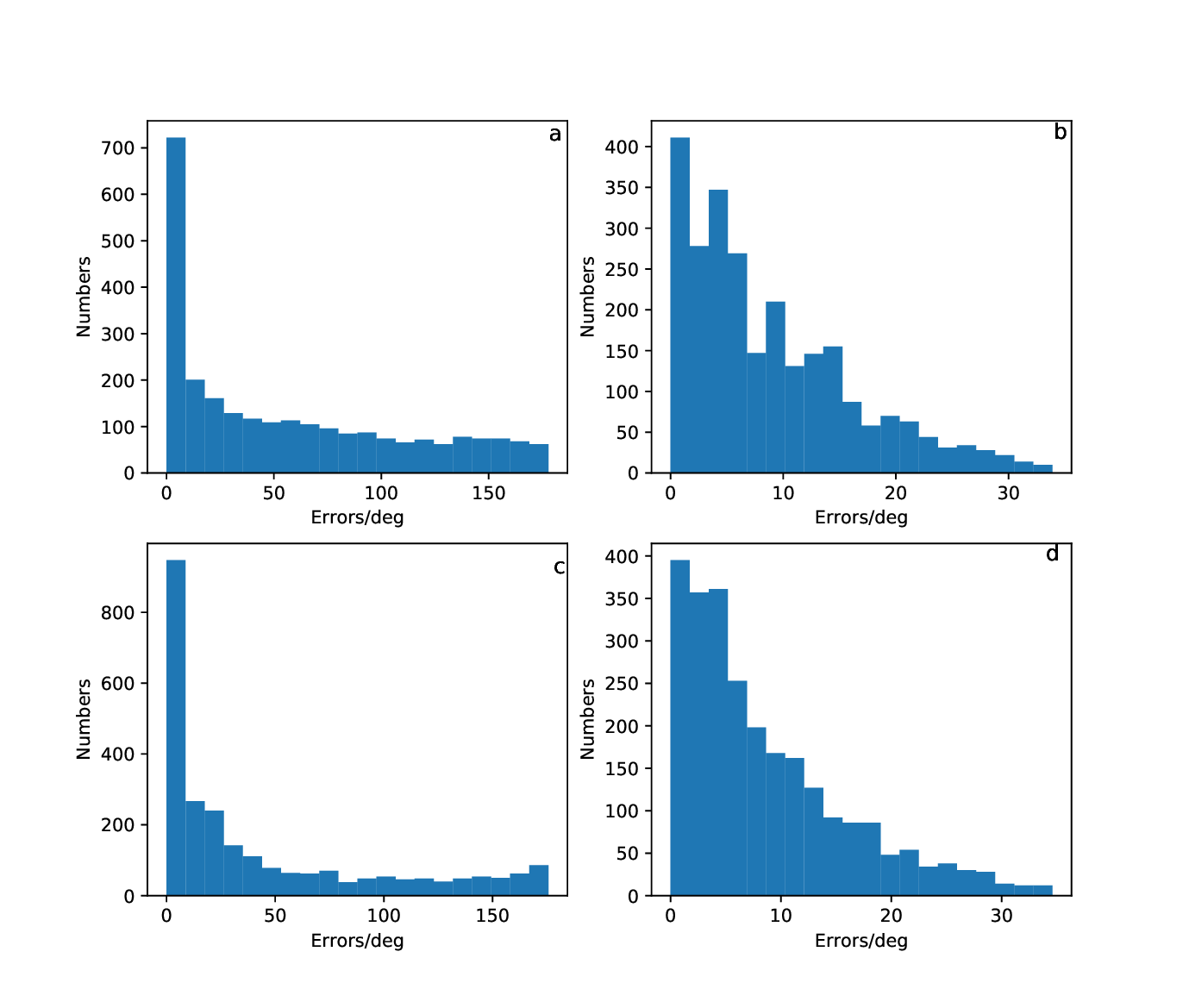}
\end{overpic}
\renewcommand{\figurename}{Fig.}
\caption{\footnotesize{a. Histogram of $\vert\lambda-\lambda_{\rm s}\vert$ when an error of $0.1\widetilde{I}$ is added to $\widetilde{I}$. b. Histogram of $\vert\bar{\theta}-\bar{\theta_{\rm s}}\vert$ when an error of $0.1\widetilde{I}$ is added to $\widetilde{I}$. c. Histogram of $\vert\lambda-\lambda_{\rm s}\vert$ when an error of $-0.1\widetilde{I}$ is added to $\widetilde{I}$. d. Histogram of $\vert\bar{\theta}-\bar{\theta_{\rm s}}\vert$ when an error of $-0.1\widetilde{I}$ is added to $\widetilde{I}$.}}
\label{fig6}
\end{figure}

\subsection{In the case of errors}

If the errors of $\Delta\widetilde{B_x}=\Delta\widetilde{B_y}=\Delta\widetilde{B_z}$ and they are proportional to $\widetilde{B_x}$, $\widetilde{B_y}$ and $\widetilde{B_z}$, they can be completely eliminated in Eq. (\ref{eq6}) or Eq. (\ref{eq7}), and the set $\lambda$ and $\bar{\theta}$ are inferred perfectly. The results are almostly the same to those in Figure \ref{fig1}.
 
However, if $\Delta\widetilde{B_x}\neq\Delta\widetilde{B_y}\neq\Delta\widetilde{B_z}$, the set $\lambda$ and $\bar{\theta}$ may not be  inferred correctly from Eq. (\ref{eq6}) or Eq. (\ref{eq7}). To investigate the errors caused by these measurement errors, here we add $\pm 10\%$ error to $\widetilde{B_x}$, $\widetilde{B_y}$, $\widetilde{B_z}$, $\widetilde{D}$, and $\widetilde{I}$, respectively. The $\widetilde{B_x}$, $\widetilde{B_y}$, $\widetilde{B_z}$, $\widetilde{D}$ and $\widetilde{I}$ for 2555 points on a longitude/latitude grid ($0.0^\circ:5.0^\circ:360.0^\circ{\rm E}, -85.0^\circ:5.0^\circ:85.0^\circ{\rm N}$) are calculated from IGRF model. Figure \ref{fig2}-\ref{fig6} show the histograms of $\vert\lambda-\lambda_{\rm s}\vert$ and $\vert\bar{\theta}-\bar{\theta_{\rm s}}\vert$ (where $\lambda_{\rm s}$ and $\bar{\theta_{\rm s}}$ are the set longitude and latitude, respectively) with these geomagnetic "measurements".  It can be seen that (See Table \ref{tb1} for details): 

(1) The uncertainty of $\lambda$ derived is obviously greater than that of $\bar{\theta}$. About 55\% of $\vert\bar{\theta} -\bar{\theta_s}\vert\leq 5^\circ$; About 68\% of $\vert\bar{\theta} -\bar{\theta_s}\vert\leq 10^\circ$; About 95\% of $\vert\bar{\theta} -\bar{\theta_s}\vert\leq 20^\circ$. However, only about 32\% of $\vert\lambda -\lambda_s\vert\leq 5^\circ$; About 42\% of $\vert\lambda -\lambda_s\vert\leq 10^\circ$; About 53\% of $\vert\lambda -\lambda_s\vert\leq 20^\circ$. And on some grid points, the errors in $\lambda$ can be up to more than $150^\circ$. 

(2) Among $\Delta\widetilde{B_x}$, $\Delta\widetilde{B_y}$, $\Delta\widetilde{B_z}$, $\Delta\widetilde{D}$ and $\Delta\widetilde{I}$, $\Delta\widetilde{I}$ causes the greatest error to $\lambda$, and only 37\% of $\vert\lambda -\lambda_s\vert\leq 20^\circ$. $\Delta\widetilde{D}$ causes the least error to $\lambda$, and about 63\% of $\vert\lambda -\lambda_s\vert\leq 20^\circ$. On the other hand, the errors by $\Delta\widetilde{D}$ to $\bar{\theta}$ are less than those by $\Delta\widetilde{I}$, and 100\% of $\vert\bar{\theta} -\bar{\theta_s}\vert\leq 20^\circ$. 

(3) The error distributions of $\lambda$ and $\bar{\theta}$ resulted from the positive or negative error perturbations are similar, but the errors from $\Delta\widetilde{I}$ and $-\Delta\widetilde{I}$ are slightly larger than those from $\Delta\widetilde{B_x}$, $\Delta\widetilde{B_y}$, $\Delta\widetilde{B_z}$, and $\Delta\widetilde{D}$. 

(4) The error distribution from $\Delta\widetilde{B_x}$ is similar to that of $\Delta\widetilde{B_z}$, but they are different from that by $\Delta\widetilde{B_y}$. $\Delta\widetilde{B_y}$ causes greater errors in $\lambda$ than those caused by $\Delta\widetilde{B_x}$ and $\Delta\widetilde{B_z}$, while $\Delta\widetilde{B_y}$ results in less errors in $\bar{\theta}$ than those resulted from $\Delta\widetilde{B_x}$ and $\Delta\widetilde{B_z}$. 

\begin{table}[htbp] 
\centering
\footnotesize
 \begin{threeparttable}
  \caption{\label{tb1}{Errors of longitude and latitude derived from Eq. (\ref{eq6}) or Eq. (\ref{eq7}) when $\Delta\widetilde{B_x}\neq\Delta\widetilde{B_y}\neq\Delta\widetilde{B_z}$.}}
  \begin{tabular}{lcccccc}
 \toprule 
 Variable (error) & $\vert\lambda-\lambda_{\rm s}\vert$ & $\vert\bar{\theta}-\bar{\theta_{\rm s}}\vert$& $\vert\lambda-\lambda_{\rm s}\vert$ & $\vert\bar{\theta}-\bar{\theta_{\rm s}}\vert$& $\vert\lambda-\lambda_{\rm s}\vert$ & $\vert\bar{\theta}-\bar{\theta_{\rm s}}\vert$ \\
 \ & $\leq 5^\circ$ & $\leq 5^\circ$& $\leq 10^\circ$ & $\leq 10^\circ$& $\leq 20^\circ$ & $\leq 20^\circ$ \\
 \midrule
$\Delta\widetilde{B_x}$ (+10\%)& 38\% &  56\% &   46\% & 66\% &  52\% &   92\%\\
$\Delta\widetilde{B_x}$(-10\%) & 39\% &  59\% &   48\% & 68\% &  58\% &   91\%\\
$\Delta\widetilde{B_y}$(+10\%) & 29\% &  61\% &   41\% & 68\% &  56\% &   98\%\\
$\Delta\widetilde{B_y}$ (-10\%)& 25\% &  53\% &   39\% & 66\% &  48\% &   97\%\\
$\Delta\widetilde{B_z}$(+10\%) & 37\% &  57\% &   45\% & 67\% &  52\% &   91\%\\
$\Delta\widetilde{B_z}$ (-10\%)& 39\% &  60\% &   49\% & 69\% &  59\% &   92\%\\
$\Delta\widetilde{D}$(+10\%)   & 32\% &  67\% &   47\% & 73\% &  63\% &   100\%\\
$\Delta\widetilde{D}$(-10\%)   & 28\% &  58\% &   41\% & 69\% &  53\% &   100\%\\
$\Delta\widetilde{I}$(+10\%)   & 23\% &  40\% &   29\% & 64\% &  37\% &   90\%\\
$\Delta\widetilde{I}$ (-10\%)  & 30\% &  43\% &   39\% & 67\% &  51\% &   91\%\\

\bottomrule 
\end{tabular} 
\tiny Notes: 

\hspace{1em}1. $\lambda$, $\bar{\theta}$ are longitude and latitude derived, respectively. $\lambda_{\rm s}$, $\bar{\theta_{\rm s}}$ are the set longitude and latitude, respectively. 

\hspace{1em}2. The $\widetilde{B_x}$, $\widetilde{B_y}$, $\widetilde{B_z}$, $\widetilde{D}$ and $\widetilde{I}$ for 2555 points on a longitude/latitude grid ($0.0^\circ:5.0^\circ:360.0^\circ{\rm E}, -85.0^\circ:5.0^\circ:85.0^\circ{\rm N}$) are calculated from IGRF model in the year 2009.

\hspace{1em}3. When an error of $\pm 0.1\widetilde{B_x}$ is added to $\widetilde{B_x}$, $\Delta\widetilde{B_y}=\Delta\widetilde{B_z}=0$. So do $\widetilde{B_y}$, $\widetilde{B_z}$, $\widetilde{D}$ and $\widetilde{I}$.

 \end{threeparttable} 
\end{table}

\begin{figure}[htb]
\setlength{\belowcaptionskip}{0pt}
\centering
\begin{overpic}[scale=0.5,bb=1 133 583 650]{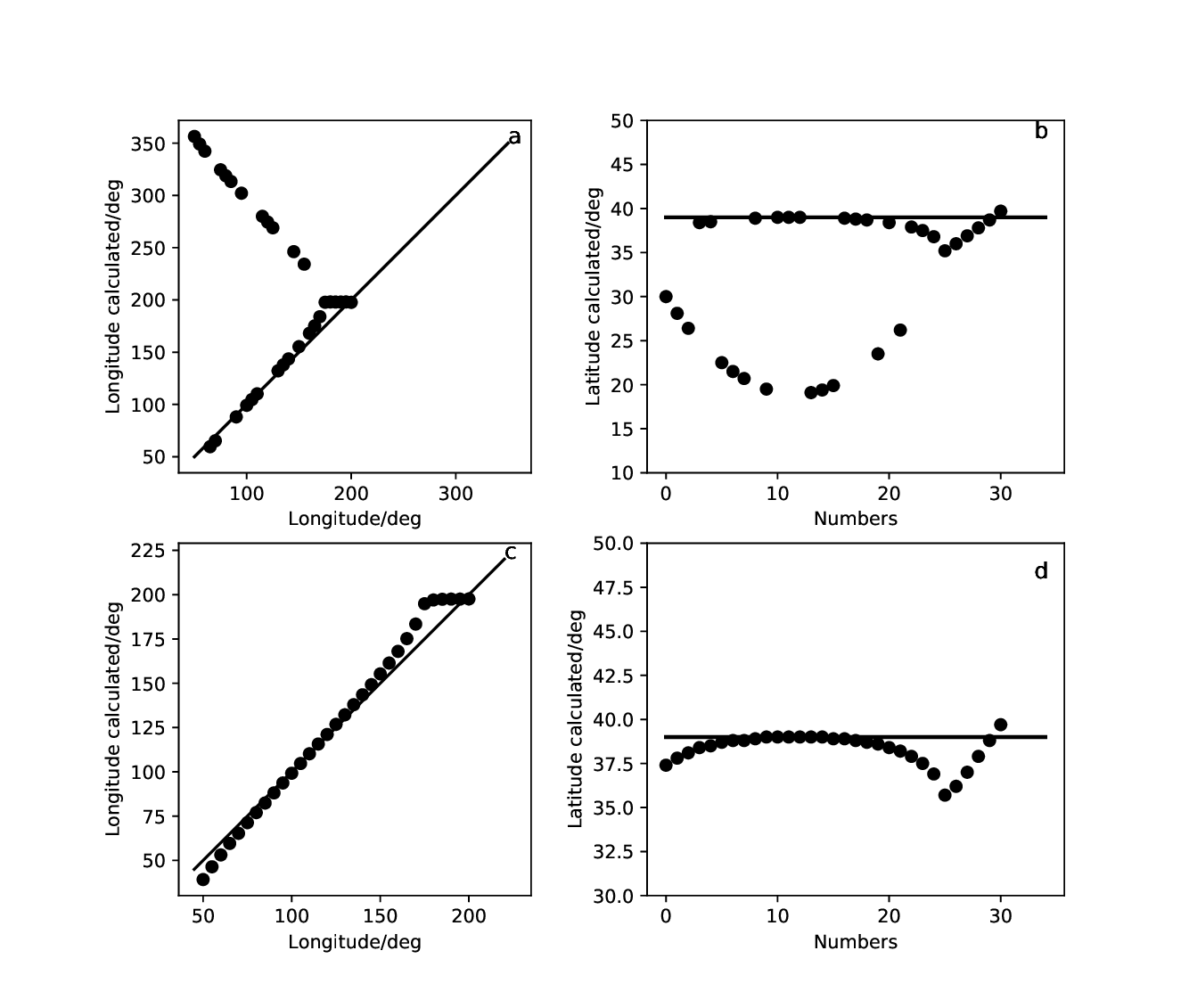}
\end{overpic}
\renewcommand{\figurename}{Fig.}
\caption{\footnotesize{a. Longitudes derived from Eq. (\ref{eq6}) vs. the set longitudes. b. Latitudes derived from Eq. (\ref{eq6}) vs. the set latitude (the black line). c. Longitudes derived from Eq. (\ref{eq10}) vs. the set longitudes. d. Latitudes derived from Eq. (\ref{eq10}) vs. the set latitude (the black line).  $\bar{\theta_s}=39^\circ$N, and and $\lambda_s$ is from $50^\circ{\rm E}$ to $200^\circ{\rm E}$ with a step of $5^\circ$.  An error of $0.1\widetilde{D}$ is added to $\widetilde{D}$.}}
\label{fig7}
\end{figure}

\begin{figure}[htb]
\setlength{\belowcaptionskip}{0pt}
\centering
\begin{overpic}[scale=0.5,bb=1 133 583 650]{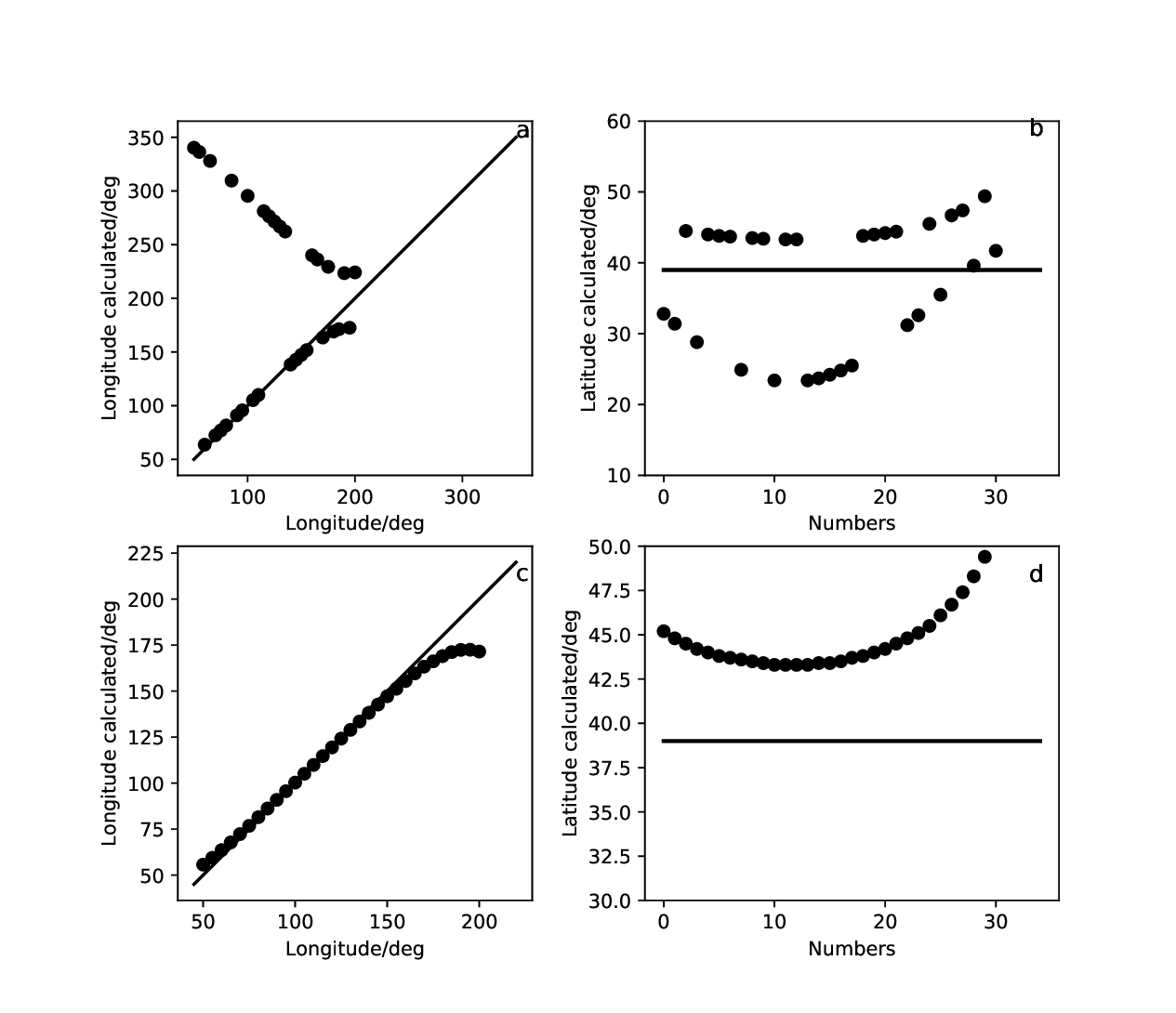}
\end{overpic}
\renewcommand{\figurename}{Fig.}
\caption{\footnotesize{a. Longitudes derived from Eq. (\ref{eq6}) vs. the set longitudes. b. Latitudes derived from Eq. (\ref{eq6}) vs. the set latitude (the black line). c. Longitudes derived from Eq. (\ref{eq10}) vs. the set longitudes. d. Latitudes derived from Eq. (\ref{eq10}) vs. the set latitude (the black line). $\bar{\theta_s}=39^\circ$N, and and $\lambda_s$ is from $50^\circ{\rm E}$ to $200^\circ{\rm E}$ with a step of $5^\circ$. An error of $0.1\widetilde{I}$ is added to $\widetilde{I}$.}}
\label{fig8}
\end{figure}

Thus it is possible to derive $\lambda$ from Eq. (\ref{eq6}) or Eq. (\ref{eq7}) with geomagnetic measurements with errors. However, the solution to $\lambda$ may be unstable as we mentioned in the previous section.  To obtain an stable and appropriate $\lambda$ (even $\bar{\theta}$), an approach like the Tikhonov regularization should be used through Eq. (\ref{eq9}) or Eq. (\ref{eq10}). It will spend more time to get $\lambda$ and $\bar{\theta}$ because of searching an appropriate regularization parameter $\alpha$.  Figure \ref{fig7} and \ref{fig8} show two examples with Eq. (\ref{eq10}), in which $\bar{\theta_s} =39^\circ{\rm N}$ and $\lambda_s$ is from $50^\circ{\rm E}$ to $200^\circ{\rm E}$ with a step of $5^\circ$. $\lambda_0=0^\circ$, and $\bar{\theta_0}=0^\circ$.  $\widetilde{D}$ and $\widetilde{I}$ are calculated from IGRF model in the year 2009. An error of $0.1\Delta\widetilde{D}$ and $0.1\Delta\widetilde{I}$ is added to $\widetilde{D}$ and $\widetilde{I}$, respectively. We systematically search through all $\alpha\in[10^{-5},10^{10}]$ with a step of $10^{0.1}$ besides all $\lambda\in [0^\circ,360^\circ]$, $\bar{\theta}\in[-90^\circ,90^\circ]$ for the local minima of the $R$ in Eq. (\ref{eq10}).

\begin{figure}[htb]
\setlength{\belowcaptionskip}{0pt}
\centering
\begin{overpic}[scale=0.5,bb=9 218 579 539]{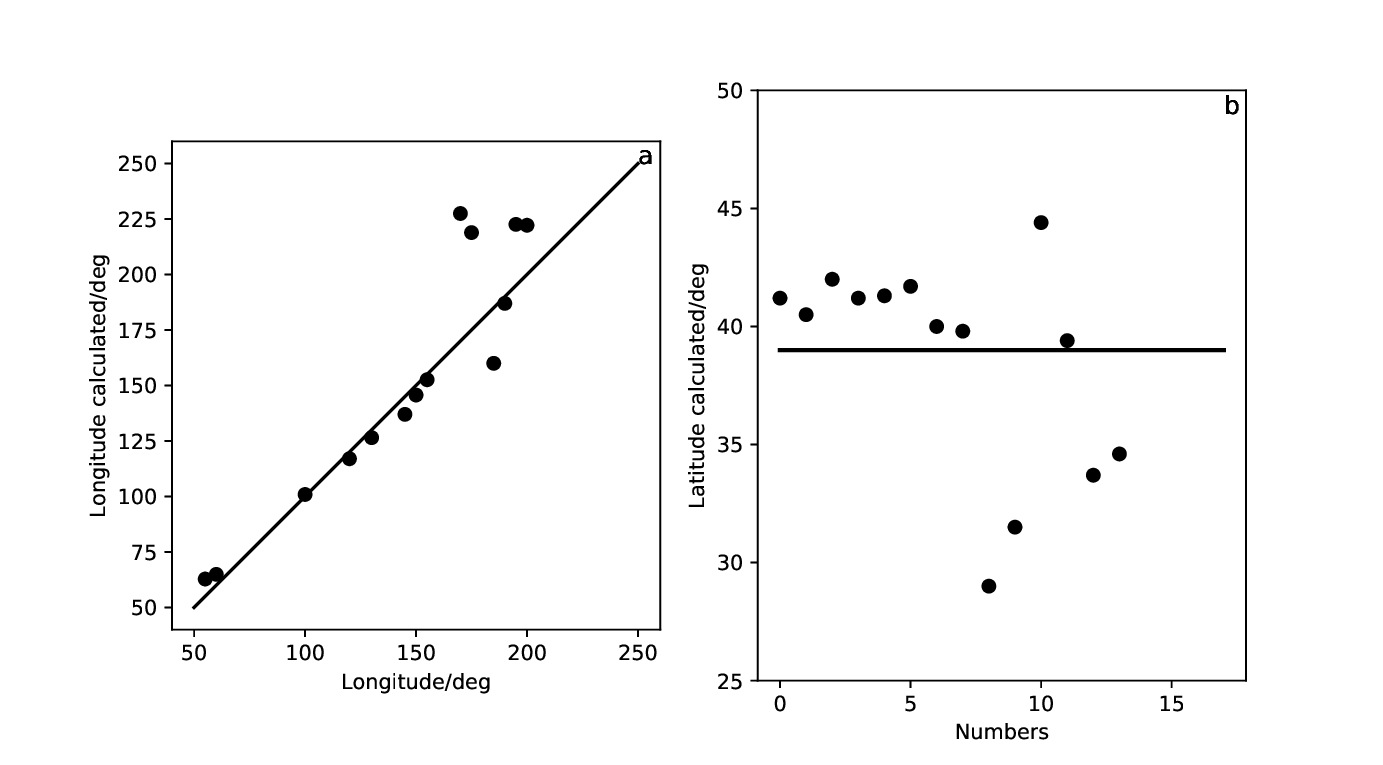}
\end{overpic}
\renewcommand{\figurename}{Fig.}
\caption{\footnotesize{a. Longitudes derived from Eq. (\ref{eq6}) vs. the set longitudes. b. Latitudes derived from Eq. (\ref{eq6}) vs. the set latitudes. Here $\lambda$, $\theta$, $\lambda_p$ and $\theta_p$ are unkonwn.}}
\label{fig9}
\end{figure}

\begin{figure}[htb]
\setlength{\belowcaptionskip}{0pt}
\centering
\begin{overpic}[scale=0.5,bb=23 223 555 532]{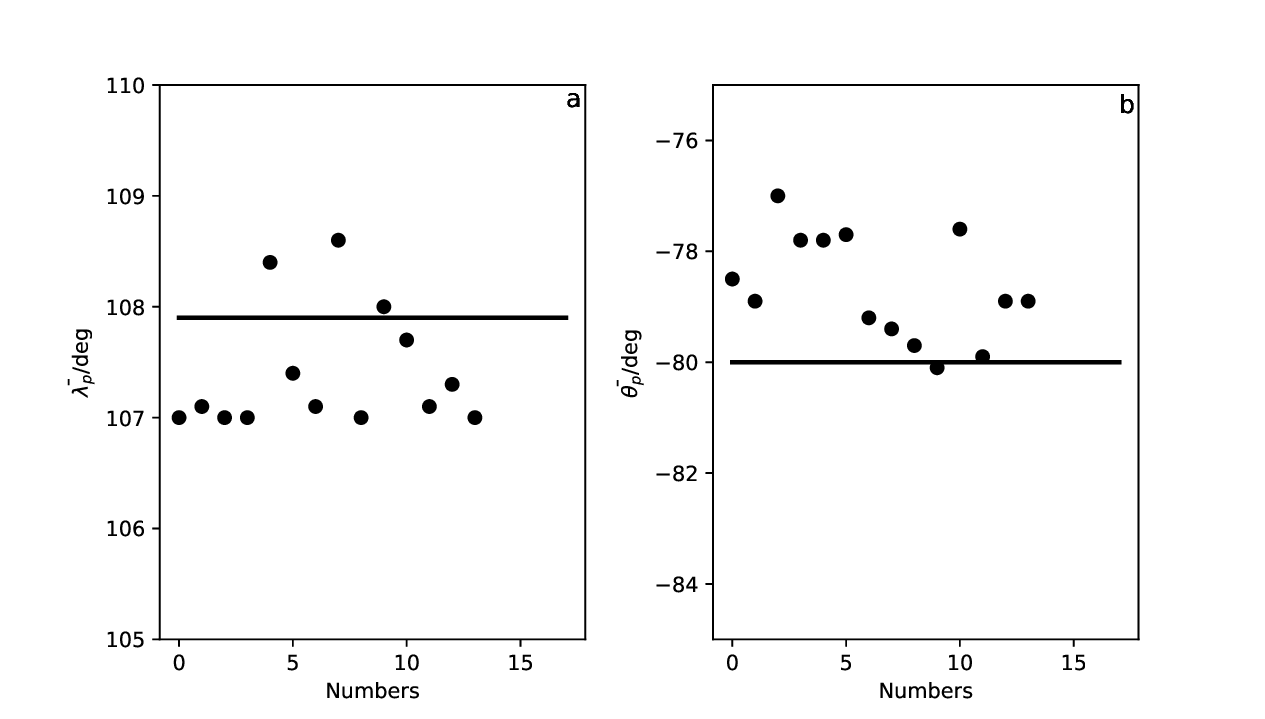}
\end{overpic}
\renewcommand{\figurename}{Fig.}
\caption{\footnotesize{a. $\lambda_p$ derived from Eq. (\ref{eq6}) vs. the true $\lambda_p$ (black solid line). b. $\theta_p$ derived from Eq. (\ref{eq6}) vs. the true $\theta_p$ (black solid line). Here $\lambda$, $\theta$, $\lambda_p$ and $\theta_p$ are unkonwn.}}
\label{fig10}
\end{figure}

Comparing Figure \ref{fig7}a with Figure \ref{fig7}c, and Figure \ref{fig8}a with Figure \ref{fig8}c, it can be seen that the accuracy of $\lambda$ derived has been greatly improved after applying the regularization method. For some grid points, $\lambda$ can be derived correctly. So does $\bar{\theta}$ if we comparing Figure \ref{fig7}b with Figure \ref{fig7}d, and Figure \ref{fig8}b with Figure \ref{fig8}d. Therefore, Eq. (\ref{eq9}) or Eq. (\ref{eq10}) can be used to infer $\lambda$ and $\bar{\theta}$ with geomagnetic measurements including errors if an appropriate regularization is obtained.

\section{Discussions}\label{sec4}

\subsection{Errors from $\lambda_p$ and $\theta_p$} In the previous part we assume that $\lambda_p$ and $\theta_p$ have no errors. This assumption should be discussed further, and we add an error of $\pm$ 10\% to $\lambda_p$ and $\theta_p$ again, respectively.  Numerical experiments on the 2555 grid points above show that: 

(1) For Eq. (\ref{eq6}), $\theta_p$ has a greater influence on $\lambda$. When $\theta_p$ has $\pm10$\% errors disturbance, about 70\% of $\vert\lambda-\lambda_s\vert \geq 20^\circ$; 

(2) For $\bar{\theta}$, when $\theta_p$ has a $+10$\% error perturbation, abount 91\% of $\vert\bar{\theta}-\bar{\theta_s}\vert\leq10^\circ$, and when there is a -10\% error perturbation, about 31\% of $\vert\bar{\theta}-\bar{\theta_s}\vert\leq 10^\circ$, but 35\% of $\vert\bar{\theta}-\bar{\theta_s}\vert \geq 20^\circ$; 

(3) When $\lambda_p$ has a 10\% error perturbation, about 89\% of $\vert\lambda-\lambda_s\vert \leq 15^\circ$, and this ratio rises to 92\% when $\lambda_p$ has a -10\% error perturbation. In addition, $\lambda_p$ has little effects on $\bar{\theta_s}$, and more than 91\% of $\vert\bar{\theta}-\bar{\theta_s}\vert\leq 1^\circ$; 

(4) $\lambda_p$ and $\theta_p$ have little influences on $\lambda$ and $\bar{\theta}$ when $\lambda$ is derived from with Eq. (\ref{eq13}) and (\ref{eq15}).

When the errors in $\lambda_p$ and $\theta_p$  are too large to be used, we still can infer $\lambda$ and $\bar{\theta}$ in theory by scanning all $\lambda$ and $\lambda_p\in [0^\circ,360^\circ]$, and all $\bar{\theta}$ and $\theta_p\in[-90^\circ,90^\circ]$ for the local minima of the $R$ in Eq. (\ref{eq6}) or Eq. (\ref{eq7}). However, numerical experiences show the finial results strongly depend on the initial value of $\lambda_p$ and $\theta_p$ because it is not only a nonlinear but also an underdetermined problem, and it will take more time.  Figure \ref{fig9} and \ref{fig10} show some "good" results. It can be seen that the $\lambda$ derived is not accurate even if there are no errors in the geomagnetic measurements. So do $\bar{\theta}$, $\lambda_p$ and $\theta_p$. When there are errors in the geomagnetic measurements, the $\lambda$ and $\bar{\theta}$ may be unstable and unreasonable and , Eq. (\ref{eq9}) or Eq. (\ref{eq10}) should be used and it will take a long time to infer $\lambda$, $\theta$, $\lambda_p$ and $\theta_p$. 

It may be a good way for constructing accurate reference poles of a certain geological age to get the reasonable initial value of $\lambda_p$ and $\theta_p$.  This demands a large amount of accurate paleo-magnetic measurements and close cooperation of scientists of paleo-magnetism. At present, there is a lot of databases can be used for this purpose,  such as IAGA paleo-magnetic databases (McElhinny and Lock, 1996; McElhinny and McFadden, 1997; McElhinny et al., 1998),  IAGA paleointensity database (Perrin et al., 1998; Perrin and Schnepp, 2004),  Absolute Palaeointensity (PINT) Database (Biggin et al., 2010; Veikkolainen et al., 2017), Magnetics Information Consortium (MagIC) (Jarboe et al., 2012), Precambrian database (PALEOMAGIA) (Veikkolainen et al., 2017), and so on. 

\subsection{In case of an known $\bar{\theta}$}  There has been a lot of work to determine the $\bar{\theta}$ based on Eq. (\ref{eq4}). And a large amount of reasonable data on $\bar{\theta}$ has been accumulated. Is it possible to use these data in the paleo-longitude study? The following experiments show that $\lambda$ (even $K_p$) can be easily obtained with the known $\widetilde{B_x}$, $\widetilde{B_y}$, $\widetilde{B_z}$, $\theta_p$, and $\lambda_p$, if $\bar{\theta}$ ($\theta=\pi/2-\bar{\theta}$) is known.
 
 From Eq. (\ref{eq5}), one can get,
 
  \begin{equation}\label{eq12}
  \left\{
  \begin{array}{ll}
 \cos(\lambda-\lambda_p)&=\frac{\cos\theta_p(\sin\theta-\frac{\widetilde{B_x}}{0.5\widetilde{B_z}}\cos\theta)}{\sin\theta_p(\cos\theta+\frac{\widetilde{B_x}}{0.5\widetilde{B_z}}\sin\theta)}\\
\sin(\lambda-\lambda_p)&=\frac{\widetilde{B_y}}{K_p\sin\theta_p}\\
K_p&=\pm\left[{\widetilde{B_x}^2+\widetilde{B_y}^2+(\frac{\widetilde{B_z}}{2})^2}\right]^\frac{1}{2}
 \end{array}
 \right.
 \end{equation}

Further, one can infer $\lambda$ from Eq. (\ref{eq12}) as the following,

\begin{equation}\label{eq13}
\lambda=\lambda_p+\tan^{-1}\left\{\pm\frac{\widetilde{B_y}\sec\theta_p}{\left[{\widetilde{B_x}^2+\widetilde{B_y}^2+(\frac{\widetilde{B_z}}{2})^2}\right]^\frac{1}{2}} \cdot\frac{\cos\theta+\frac{\widetilde{B_x}}{0.5\widetilde{B_z}}\sin\theta}{\sin\theta-\frac{\widetilde{B_x}}{0.5\widetilde{B_z}}\cos\theta}\right\}
\end{equation}

One can also get $\lambda$ with the known declination ($\widetilde{D}$) and inclination($\widetilde{I}$), $\theta_p$, and $\lambda_p$, because the following equations hold (e.g., Kono and Tanaka, 1995).

\begin{equation}\label{eq14}
\left\{
\begin{array}{ll}
\frac{\widetilde{B_y}}{\widetilde{B_x}}&=\tan \widetilde{D}\\
\frac{\widetilde{B_x}}{0.5\widetilde{B_z}}&=2\cot \widetilde{I}\cos \widetilde{D}\\
\tan\widetilde{I}&=2\cot\theta\\
\sin\theta&=\pm\left[\frac{\widetilde{B_x}^2+\widetilde{B_y}^2}{\widetilde{B_x}^2+\widetilde{B_y}^2+(\widetilde{B_z}/2)^2}\right]^{1/2}
\end{array}
\right.
\end{equation}

From Eq. (\ref{eq12}) and (\ref{eq13}), one can obtain,

\begin{equation}\label{eq15}
\lambda=\lambda_p+\tan^{-1}\left[\pm\frac{\sin\widetilde{D}\sec\theta_p}{\left(\frac{\tan^2 \widetilde{I}}{4}+1\right)^{1/2}}\cdot\frac{\cos\theta+2\cos \widetilde{D} \cot \widetilde{I} \sin\theta}{\sin\theta-2\cos \widetilde{D} \cot \widetilde{I}\cos\theta}\right]
\end{equation}

Explicit Eq. (\ref{eq13}) and (\ref{eq15}) can be used to infer the $\lambda$ when 
$\theta$, $\theta_p$, and $\lambda_p$ are known and $\widetilde{B_x}$, $\widetilde{B_y}$, $\widetilde{B_z}$ (or $\widetilde{D}$, $\widetilde{I}$) are measured. When there are no errors in the geomagnetic "measurements" (eg., $\widetilde{B_x}$), our synthetic experiments show that the $\lambda$ can be derived correctly. 

When there are errors in the geomagnetic "measurements", the unreasonable $\lambda$ may be derived. We investigate the errors caused through deriving $\lambda$ on the grid points in the previous section. We still add an error of $\pm$ 10\% to $\widetilde{B_x}$,$\widetilde{B_y}$,$\widetilde{B_z}$, $\widetilde{D}$, $\widetilde{I}$ and $\bar{\theta}$, respectively.  The results are shown in Table \ref{tb2}. 

\begin{table}[htbp] 
\centering
\footnotesize
 \begin{threeparttable}
  \caption{\label{tb2}{Errors of longitude derived from Eq. (\ref{eq13}) or Eq. (\ref{eq15}) when there are geo-magnetic measurement errors and $\bar{\theta}$.}}
  \begin{tabular}{lccc}
 \toprule 
 Variable (error) & $\vert\lambda-\lambda_{\rm s}\vert$ & $\vert\lambda-\lambda_{\rm s}\vert$ & $\vert\lambda-\lambda_{\rm s}\vert$ \\
 \ & $\leq 5^\circ$ & $\leq 10^\circ$ & $\leq 20^\circ$ \\
 \midrule
$\Delta\widetilde{B_x}$ (+10\%)& 43\% &  72\% &   100\% \\
$\Delta\widetilde{B_x}$(-10\%) & 41\% &  68\% &   100\% \\
$\Delta\widetilde{B_y}$(+10\%) & 100\% &  - &   - \\
$\Delta\widetilde{B_y}$ (-10\%)& 100\% &  - &   - \\
$\Delta\widetilde{B_z}$(+10\%) & 43\% &  72\% &   100\% \\
$\Delta\widetilde{B_z}$ (-10\%)& 41\% &  68\% &   100\% \\
$\Delta\widetilde{D}$(+10\%)   & 96\% &  100\% &   - \\
$\Delta\widetilde{D}$(-10\%)   & 97\% &  100\% &   - \\
$\Delta\widetilde{I}$(+10\%)   & 21\% &  32\% &   48\% \\
$\Delta\widetilde{I}$ (-10\%)  & 22\% &  36\% &   57\% \\
$\bar{\theta}$(+10\%)   & 24\% &  41\% &   65\% \\
$\bar{\theta}$ (-10\%)  & 24\% &  41\% &   65\% \\
\bottomrule 
\end{tabular} 
\tiny Notes: 

\hspace{1em}1. $\lambda$ and $\lambda_{\rm s}$ are longitudes derived and the set, respectively.  

\hspace{1em}2. The $\widetilde{B_x}$, $\widetilde{B_y}$, $\widetilde{B_z}$, $\widetilde{D}$ and $\widetilde{I}$ for 2555 points on a longitude/latitude grid ($0.0^\circ:5.0^\circ:360.0^\circ{\rm E}, -85.0^\circ:5.0^\circ:85.0^\circ{\rm N}$) are calculated from IGRF model in the year 2009.

\hspace{1em}3. When an error of $\pm 0.1\widetilde{B_x}$ is added to $\widetilde{B_x}$, $\Delta\widetilde{B_y}=\Delta\widetilde{B_z}=0$. So do $\widetilde{B_y}$, $\widetilde{B_z}$, $\widetilde{D}$ and $\widetilde{I}$.

 \end{threeparttable} 
\end{table}

From Table \ref{tb2}, it is found that the errors caused by the perturbation of $\widetilde{I}$ and $\bar{\theta}$ are the largest. Only about 23\% of $\vert\lambda -\lambda_s\vert\leq 5^\circ$; About 38\% of $\vert\lambda -\lambda_s\vert\leq 10^\circ$. But if $\widetilde{I}$ and $\bar{\theta}$ are accurate enough, the errors caused by $\widetilde{B_x}$, $\widetilde{B_z}$, $\widetilde{B_y}$, and $\widetilde{D}$ are small, and all $\vert\lambda -\lambda_s\vert\leq 20^\circ$ for these grid points. Especially all $\vert\lambda -\lambda_s\vert\leq 5^\circ$ when the error is only from $\widetilde{B_y}$. The above shows that if $\widetilde{I}$ and/or $\bar{\theta}$ are determined well, we can infer the $\lambda$ with Eq. (\ref{eq13}) and (\ref{eq15}) easily.

\subsection{Other cost functions}
The cost function of Eq. (\ref{eq6}) or Eq. (\ref{eq7}) is not unique, and there are other cost functions. For example,

\begin{equation}\label{eq16}
R=\left(\frac{\widetilde{B_y}}{\widetilde{B_x}}-\frac{B_y}{B_x}\right)^2+\left(\frac{\widetilde{B_y}}{\widetilde{B_z}}-\frac{B_y}{B_z}\right)^2+\left(\frac{\widetilde{B_z}}{\widetilde{B_x}}-\frac{B_z}{B_x}\right)^2
\end{equation}

By minimizing the cost function of Eq. (\ref{eq16}) with scanning method, we also can get $\lambda$ and $\bar{\theta}$ simultaneously.

\section{Conclusions}\label{sec5}

Assuming that $\theta_p$ and $\lambda_p$ are determined well, we define a cost function $R$ in Eq. (\ref{eq6}) or Eq. (\ref{eq7}), from which the paleo-longitude $\lambda$ and paleo-latitude $\bar{\theta}$ can be derived simultaneously with the scanning method from usual paleo-geomagnetic measurements without errors. Otherwise, an approach like the Tikhonov regularization can be used for deriving reasonable $\lambda$ and $\bar{\theta}$ simultaneously through Eq. (\ref{eq9}) or Eq. (\ref{eq10}) with the scanning method. 

Among the error of paleo-geomagnetic measurements of $\Delta\widetilde{B_x}$, $\Delta\widetilde{B_y}$ and $\Delta\widetilde{B_z}$, $\Delta\widetilde{B_y}$ causes the greast error to $\lambda$. So does $\Delta\widetilde{I}$ among $\Delta\widetilde{D}$ and $\Delta\widetilde{I}$.

When $\theta_p$ and $\lambda_p$ are not determined well and can not be used, we still can derive directly the $\lambda$ and $\bar{\theta}$ through the equations above with the scanning method in theory, although the finial results are strongly dependent on the initial values of $\theta_p$ and $\lambda_p$. In this sense we suggest a theoretical framework that can directly invert the $\lambda$ from the paleo-magnetic measurements with any reasonable inversion method, and invert the $\bar{\theta}$ and even the location of the paleo-magnetic poles simultaneously.

\vspace{10em}

\def\thebibliography#1{
{\Large\bf  References}\list
 {}{\setlength\labelwidth{1.4em}\leftmargin\labelwidth
 \setlength\parsep{0pt}\setlength\itemsep{.3\baselineskip}
 \setlength{\itemindent}{-\leftmargin}
 \usecounter{enumi}}
 \def\newblock{\hskip .11em plus .33em minus -.07em}
 \sloppy
 \sfcode`\.=1000\relax}
\let\endthebibliography=\endlist

\end{document}